\documentclass[aps,onecolumn,amssymb]{revtex4}
\usepackage{epsfig,amsmath}
\usepackage{graphicx}
\usepackage{subfigure} 

\begin{document}

\title{Jamming graphs: A local approach to global mechanical rigidity}

\author{Jorge H. Lopez$^1$, L. Cao$^1$, J. M. Schwarz$^1$}
\affiliation{$^1$Physics Department, Syracuse University, Syracuse, NY 13244}
\date{\today}
\begin{abstract}
We revisit the concept of minimal rigidity as applied to soft repulsive, frictionless sphere packings in two-dimensions with the introduction of the jamming graph. Minimal rigidity is a purely combinatorial property encoded via Laman's theorem in two-dimensions. It constrains the global, average coordination number of the graph, for example. However, minimal rigidity does not address the geometry of local mechanical stability. The jamming graph contains both properties of global mechanical stability at the onset of jamming and local mechanical stability. We demonstrate how jamming graphs can be constructed using local moves via the Henneberg construction such that these graphs fall under the jurisdiction of correlated percolation.  We then probe how jamming graphs destabilize, or become unjammed, by deleting a bond and computing the resulting rigid cluster distribution. We also study how the system restabilizes with the addition of new contacts and how a jamming graph with extra/redundant contacts destabilizes.  The latter endeavor allows us to probe a disc packing in the rigid phase and uncover a potentially new diverging lengthscale associated with the random deletion of contacts as compared to the study of cut-out (or frozen in) subsystems.    
\end{abstract}
\maketitle

\section{Introduction}

A model system for understanding the onset of rigidity in disordered particle packings is a $d$-dimensional collection of soft repulsive, frictionless spheres at zero temperature ~\cite{ohern1, ohern2, jamming1, jamming2, jamming3}.  At small packing densities, minimal energy configurations are those with no contacts between particles. As the packing density is increased, the contact geometry abruptly changes from nonexistent to one where the average coordination number equals $2d$ ~\cite{ohern2}. Moreover, the local coordination number for each particle must be at least $d+1$ in accordance with the Hilbert stability condition, or local mechanical stability~\cite{alexander}. As the packing density is increased even further, the average coordination number exceeds $2d$ with each particle still obeying the Hilbert stability criterion. 

The abrupt change in contact geometry indeed has the flavor of a phase transition even though, for example, no symmetries in the positions of the particles are broken ~\cite{ohern2, jamming1, jamming2}. Continuous phase transitions, such as uncorrelated percolation, are typically accompanied by at least one diverging lengthscale on either side of the transition ~\cite{percolation}. While the abrupt change in contact geometry in the soft repulsive, frictionless sphere system suggests a more exotic mixed transition, such as the found in $k$-core percolation~\cite{schwarz1}, Wyart and collaborators~\cite{wyart1,wyart2} identified a lengthscale in the disordered solid phase, denoted as $l^*$. This lengthscale can be determined by finding a cut-out subsystem of some length below which there exists extended zero energy modes within the subsystem and above which there does not. 

At the transition, extended zero-energy modes proliferate due to the absence of redundant contacts, and $l^*$ is of order the system size. As the system solidifies further with added redundant contacts, $l^*$ decreases since such contacts make it less likely for zero-energy modes to extend across the subsystem. In practice, this lengthscale $l^*$ is inferred from numerical measurements of the frequency at which the density of states deviates from the plateau region emerging at low frequencies. Very recently however, a new construction of $l^*$ via rigid clusters results in a direct numerical measurement~\cite{goodrich}. Another lengthscale associated with subsystems with fixed boundaries (as opposed to cut/free boundaries) has been recently identified and scales similarly with $l^*$~\cite{mailman1,mailman2,tighe1}, though corrections to scaling are discussed~\cite{mailman1,mailman2}. 

While the focus on identifying a diverging lengthscale has been on the competition between bulk and boundary effects when cutting out a subsystem and probing for extended zero energy modes, little work has been done to search for a lengthscale associated with a change in the zero-energy mode structure in response to the breaking of one or several contacts, for example. After all, when searching for diverging lengthscales in systems undergoing phase transitions, one typically perturbs the system and computes the lengthscale over which the system responds to that perturbation. And while prior focus on cut-out subsystems has certainly proved useful, we ask what can be learned from removing one or two contacts and looking for a lengthscale over which the perturbation affects the mode structure? Point perturbations have been performed when investigating the force network of the particle packing above jamming~\cite{ellenbroek1,ellenbroek2} and on phonon modes in floppy networks, i.e. below jamming~\cite{during}. Here, we explore random bond deletion effects on the structure of zero-energy modes via the study of rigid clusters and ask does some diverging lengthscale fall out of such computations? If so, is this lengthscale similar to $l^*$, or not? 

To begin to answer such questions, we first present a novel way to build the contact geometry of soft repulsive, frictionless discs at the onset of rigidity. We do so with vertices representing particles and bonds representing contacts between particles.  After all, at the transition there are no forces, i.e. the particles are in contact but not overlapping, so one does not necessarily need to rely on forces explicitly to generate the packing. This jamming graph algorithm uses spatially local rules to generate the contact geometry---local rules that encode the global rule of minimal rigidity. Interestingly, the uncorrelated percolation transition, with its local rules, can be described by a field theory~\cite{fieldtheory}. If the abrupt changes in contact geometry in the soft repulsive, frictionless sphere system also be characterized by local rules, can this correlated percolation system also be described by a field theory? After introducing our algorithm, we then perturb the contact geometry of the jamming graph and study the resulting rigid cluster distributions. 

The paper is organized as follows. Section II discusses the local and global properties of the contact geometry of frictionless, repulsive soft spheres. Section III presents the algorithm for building a jamming graph, Section IV addresses perturbations of the jamming graph, and Section V discusses the implications of our results.

\section{Contact geometry of frictionless, repulsive soft spheres at the onset of rigidity}

The initial link between contact geometry and the onset of rigidity in mechanical networks with fixed connectivity is due to Maxwell via the Maxwell constraint counting condition~\cite{maxwell}. This condition is a necessary (but not sufficient) for mechanical rigidity.  It does so by counting the number of zero-energy (floppy) modes in the network, $N_f$, which depends on the number of independent
constraints, $N_c$, and the local degrees of freedom for each particle. For mechanical networks with particles interacting via central forces, 
\begin{equation}
N_f=Nd-N_c,
\end{equation}
where $N$ is the number of particles in the network. The onset of rigidity, or minimal rigidity, occurs when $N_f$ equals the number of global degrees of freedom of the network, $N_g$. When $N_f=N_g$, the network is minimally rigid and the removal of just 
one edge in the network makes it flexible. In mean-field, one can replace $N_c$ with $<z>\frac{N}{2}$, where
$<z>$ is the average coordination number such that the minimally rigid condition in the large $N$ limit becomes $<z>=2d$, i.e. isostaticity. Numerical simulations indicate that the onset of mechanical rigidity for soft repulsive, frictionless spheres corresponds to the isostatic condition~\cite{ohern2}.

\begin{figure}[h]
\centering
\includegraphics[width=5cm]{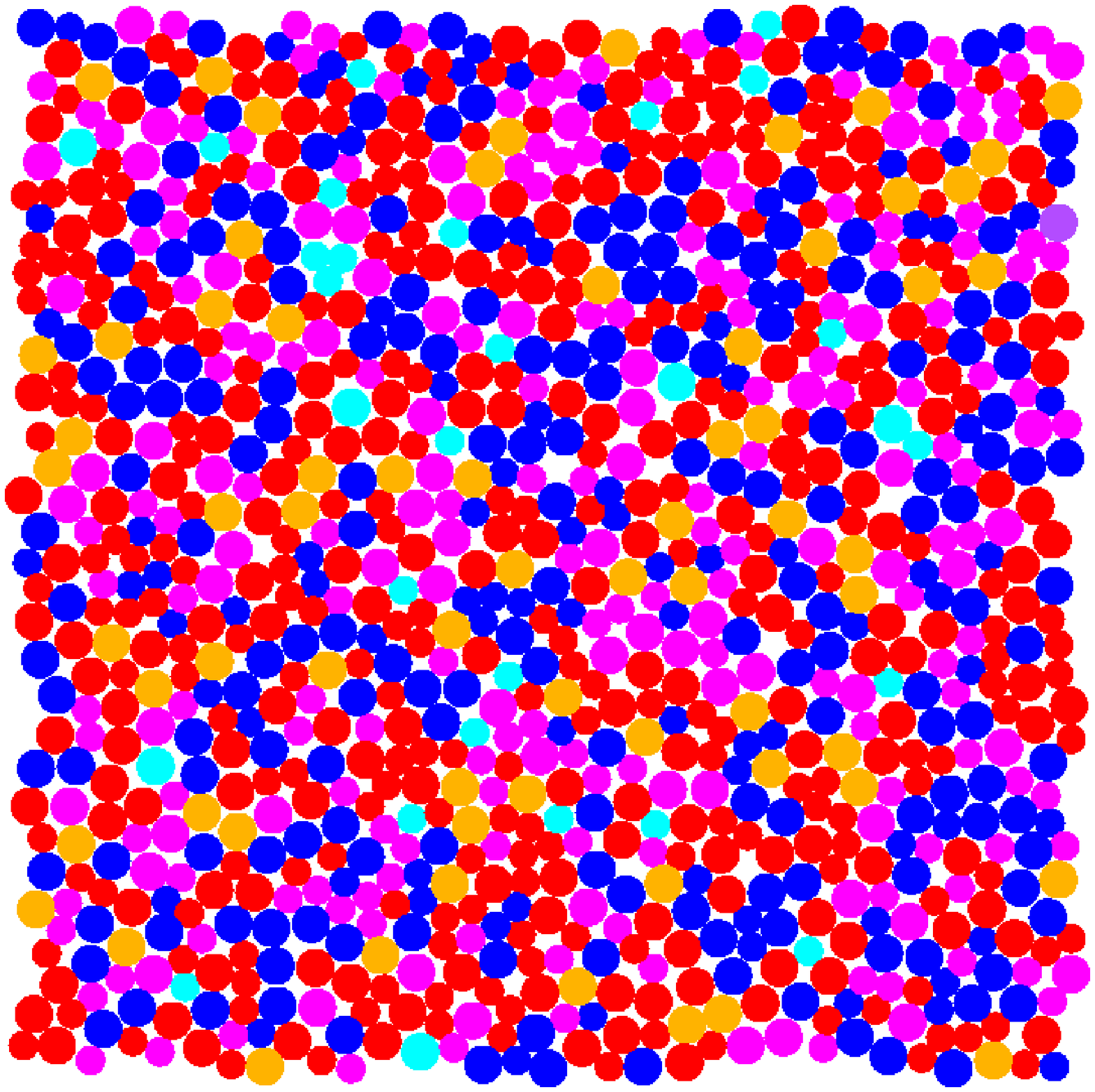}
\vspace{-2cm}
\caption{A jammed, bidisperse particle packing with $N=1024$ and
packing fraction $\phi=0.841$. The colors denote the local
coordination number, $z$, with light blue denoting $z=0$, magenta
denoting $z=3$, red denoting $z=4$, blue denoting $z=5$, orange
depicting $z=6$, and purple denoting $z=7$, which is possible for
the 1.4 diameter ratio for the bidisperse system.}
\label{fig:jammed.packing}
\end{figure}

In two-dimensions, one can extend the above necessary condition for minimal rigidity in central-force fixed connectivity networks to necessary and sufficient using Laman's theorem~\cite{laman}: {\it A network with $N$ vertices
is generically, minimally rigid in two dimensions if and only if
it has $2N-3$ bonds and no subgraph of $n$ vertices has more than
$2n-3$ bonds.}  Laman's theorem is global (or spatially nonlocal) in the sense that it involves all
possible subgraphs. However, we will present a construction of the 
Laman graph implemented via spatially local rules involving both the
addition and deletion of bonds developed earlier by Henneberg~\cite{henneberg}.
Such an algorithm falls under the jurisdiction of correlated percolation
where there are constraints on the occupation of bonds as the graph is constructed. Note that for Laman's theorem, $N_g=3$. Moreover, a recent generalization of Laman's theorem to $kN-N_g$ extends the concept of minimal rigidity ~\cite{ileana}.  

Fixed connectivity central-force networks and soft repulsive, repulsive spheres differ from each other in the following
way. In the particulate system, the contacts are not
fixed. These contacts rearrange as the system minimizes its energy, for example. 
Does this difference
have any implications for characterizing the contact geometry of the particulate system at the onset of rigidity, or jamming?  Indeed, it does. At the transition,
in addition to the system being minimally rigid, each particle
must be locally mechanical stable. Otherwise, an infinitesimal
perturbation can break a contact and the system becomes flexible.

What does local mechanical stability mean in terms of constraints on the contact geometry?
In two dimensions, a particle must have at least three contacts and those contacts must be organized in
such a way the particle cannot escape its local environment via a
perturbation. In other words, the contact particles must be
counter-balancing. If this local condition is not obeyed, then the
entire packing can go unstable to due to an infinitesimal, local
perturbation. 

Therefore, just as S. Alexander~\cite{alexander} argued that the concept of
geometrical rigidity for spring networks, when extended to
particle networks, needs to incorporate the breaking of contacts, the rigidity transition for soft repulsive, frictionless spheres in two-dimensions is characterized by a spanning, planar graph obeying both the Laman condition {\it and} by the principle of local
mechanical stability (Hilbert stability). We call this graph a ``jamming graph''.  The graph must be planar since the bonds represent particle contacts. 

We would like to understand how such graphs obeying both global and local rules of mechanical stability are constructed in practice. It turns out that we will be able to do so using an algorithm with spatially local moves. This is because one can build a Laman graph via an algorithm called the
Henneberg construction~\cite{henneberg}. The algorithm will be presented in the following section. Therefore, we now explore the notion of minimal rigidity in particle packings in
two-dimensions, where the connectivity is not fixed, by imposing {\it both} types of mechanical stability. 

\section{Algorithm for building jamming graphs}

\subsection{Henneberg construction}

\begin{figure}[h]
\centering
\includegraphics[width=12cm, bb = 146 196 620 413]{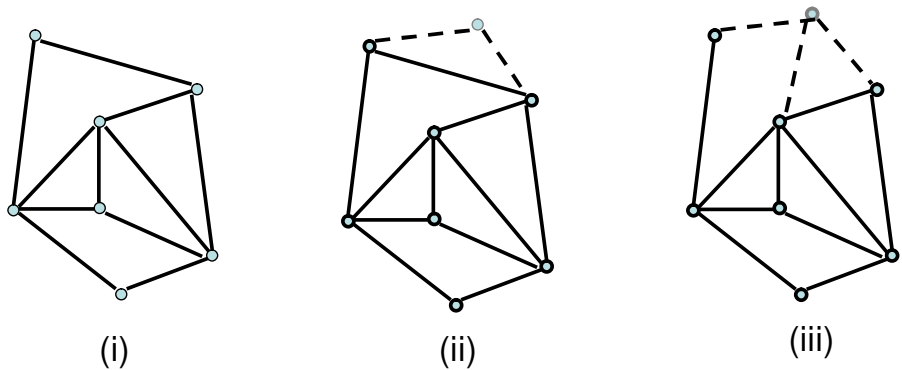}
\caption{(i) Graph before Henneberg step. (ii) Type I step with the new bonds denoted with dashed lines. (iii)
Type II step.} \label{fig:Henneberg.constructions}
\end{figure}

Let us review the Henneberg construction~\cite{henneberg}. For constructing a Laman graph, we begin with a triangle and then add a vertex and connect it to prior vertices using the Henneberg steps Type I and Type II defined below: 
\begin{itemize}
    \item Type I step: Add a vertex and join it to two prior vertices via two
    new bonds;
    \item Type II step: Add a vertex and join it to three prior vertices with
    at least one bond in between the three bonds. Remove a prior bond between the three connecting prior vertices.
\end{itemize}
See Fig.~\ref{fig:Henneberg.constructions} for an illustration of the Henneberg construction. 

A graph constructed using the Henneberg
construction is Laman~\cite{laman}. We can see this via induction. Suppose the current graph
$G$ is Laman with $N$ vertices and $2N-3$ bonds.\\
For the Type I step: Add vertex $x$. Graph $G$ now contains $N+1$ vertices and
$2N-3+2=2(N+1)-3$ bonds. For any subgraph with $n$ vertices, if it
does not include $x$, by the induction hypothesis, there are at
most $2n-3$ bonds. If the subgraph includes $x$, for the other
$n-1$ vertices, there are at most $2(n-1)-3$ bonds between them,
so in total there are at most $2(n-1)-3+2=2n-3$ bonds.\\
\noindent For the Type II step: Add vertex $x$ and connect it to $a$, $b$, $c$.
Graph $G$ now contains $N+1$ vertices and $2N-3+3-1=2(N+1)-3$ bonds. For any
subgraph with $n$ vertices, if it does not include $x$, by the
induction hypothesis, there are at most $2n-3$ bonds. If any 
subgraph includes $x$, for the other $n-1$ nodes, there are at most
    \begin{enumerate}
        \item $2(n-1)-3$ bonds, if not all of $a$, $b$, $c$ are included.
        \item $2(n-1)-4$ bonds, if $a$, $b$, $c$ are all included.
    \end{enumerate}
In case 1, there are at most $2(n-1)-3+2=2n-3$ bonds. In case 2,
there are at most $2(n-1)-4+3=2n-3$ bonds. Thus, the Laman condition is
satisfied.  One can also show that every Laman graph can be decomposed into a Henneburg construction. 

The Henneberg construction (and the corresponding Laman's theorem) is purely combinatorial.  It only deals with adjacency and not where the neighbors are located spatially. Because bonds in the jamming graph represent contacts between particles, we impose a planarity, or no-crossing condition, on the bonds. Moreover, if $N_g=2$, as opposed to $N_g=3$, the above Henneberg construction is unchanged.

\subsection{Hilbert stability}

\begin{figure}[h]
\centering
\includegraphics[width=6cm]{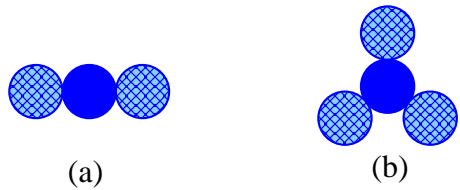}
\caption{(a) The center particle is not mechanically stable (for
the other two particles fixed). (b) The center particle is
mechanically stable.} \label{fig:stability}
\end{figure}

We also invoke the local Hilbert stability condition, otherwise known as the counter-balance condition. It states that any vertex should have at least three neighbors, i.e. it is $3-core$, and the 
vertex should be inside the triangle determined by at least three of its neighbors. Soft repulsive, frictionless particles satisfying
Hilbert stability condition are consequently locally mechanically stable. This condition is illustrated in Fig.~\ref{fig:stability}. 

\begin{figure}[h]
\centering
\includegraphics[width=4cm]{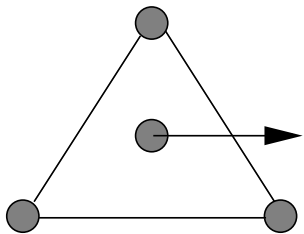}
\caption{Illustration for checking if a particle enclosed by a triangle. } \label{fig:point.in.polygon}
\end{figure}

To determine whether or not a vertex is enclosed in
a triangle by at least three of its neighbors, we implement an algorithm based on the Jordan Curve
Theorem~\cite{jordan}. It consists of drawing a horizontal line
from the vertex and determining how many crossings the horizontal
line makes with the enclosed triangle.  If there are an odd number
of crossings, then the vertex is inside the polygon as illustrated in
Fig.~\ref{fig:point.in.polygon}.

\subsection{Pseudocode}

Graphs built by using Henneberg type II steps have typically more counter-balanced vertices since the new vertex has more neighbors in comparison to Type I. So we build a jamming graph implementing Henneberg type II steps only. The graph must be planar and we implement periodic boundary conditions on a square of length unity. We then enforce
the Hilbert stability condition on each vertex. Here is the pseudocode of our algorithm. 

\begin{itemize} 
  \item Create a triangle.  The length of the bonds is chosen randomly from the uniform distribution on the interval $[r_{min},r_{max}]$, where $r_{min}<r_{max}<0.5$. The triangle is a minimally rigid $k=3$, $N_g=3$ structure.   
\item Create a new vertex with random coordinates with the constraint that no vertex be less than a distance $r_{min}$ or greater than a distance $r_{max}$ from any other vertex and establish its neighbors according to the Type II Henneberg step. This creation is succesful if the new bonds are not overlapping any of the surviving prior bonds (keeping in mind that a Type II Henneberg move implies the deletion of a prior bond). 
\item Repeat the above step $N-4$ times to create a planar Laman graph with $N$ vertices.
\item Check for those vertices that are not counter-balanced.
\item For each vertex that is not yet counter-balanced, impose the following set of strategies to enforce counter-balance. The strategies differ depending on whether an uncounter-balanced vertex has at least two neighbors or at least three. 
\item \textbf{Counter-balance strategies for a vertex with at least three neighbors}: Suppose uncounter-balanced 
		vertex $p$ has  neighbors $n_{1}$, $n_{2}$, and $n_{3}$, as shown in Figure \ref{CBpseudocode1}. If $p$ has more than three neighbors, we apply the following strategy to each set of the three diferent neighbors of $p$. Choose $n_{2}$ such that $\overrightarrow{pn_{2}}$ is inside the angle $\widehat{n_{1}pn_{3}}$. Then choose a vertex $x$ such that
	\begin{equation} 
		\overrightarrow{x}=\overrightarrow{n}+\frac{\overrightarrow{np}}{\|\overrightarrow{pn}\|}\alpha;\quad
		\overrightarrow{n}=\beta\overrightarrow{n_{1}}+\left(1-\beta\right)\overrightarrow{n_{2}},
	\end{equation}
where $\alpha$ is equal to one half the distance from $p$ to its closest neighbor and $0<\beta<1$. To counterbalance $p$ so that it belongs to the triangle 
		$\triangle n_{1}n_{2}x$, $x$ should be in the angle determined by the vectors $\overrightarrow{n_{1}p}$ and $\overrightarrow{n_{2}p}$. By choosing particular values for $\beta$, for example $\beta=\frac{1}{10},\frac{1}{2},\frac{9}{10}$, most of the region containing the angle is scanned as illustrated in Figure 
\ref{CBpseudocode1}.

New bonds $\overline{xp}$, $\overline{xn_{3}}$, $\overline{xz}$ are created and the prior bond $\overline{pn_{3}}$ is deleted. Note that the new bonds cannot cross any of the prior ones. Then, a prior vertex $z$ is chosen such that $x$ is counter-balanced, which means it should be in the striped region in Figure 
\ref{CBpseudocode1}. If the above counter-balance strategy does not work, move $p$ towards $n_{2}$ by a fraction of the length $\|\overrightarrow{pn_{2}}\|$. Once $p$ is moved, the above strategy is tried again. If this particular construction does not work, chose $n$ to be between $n_{2}$ and $n_{3}$.

\begin{figure}
\centering
\begin{tabular}{cc}
\includegraphics[width=3in]{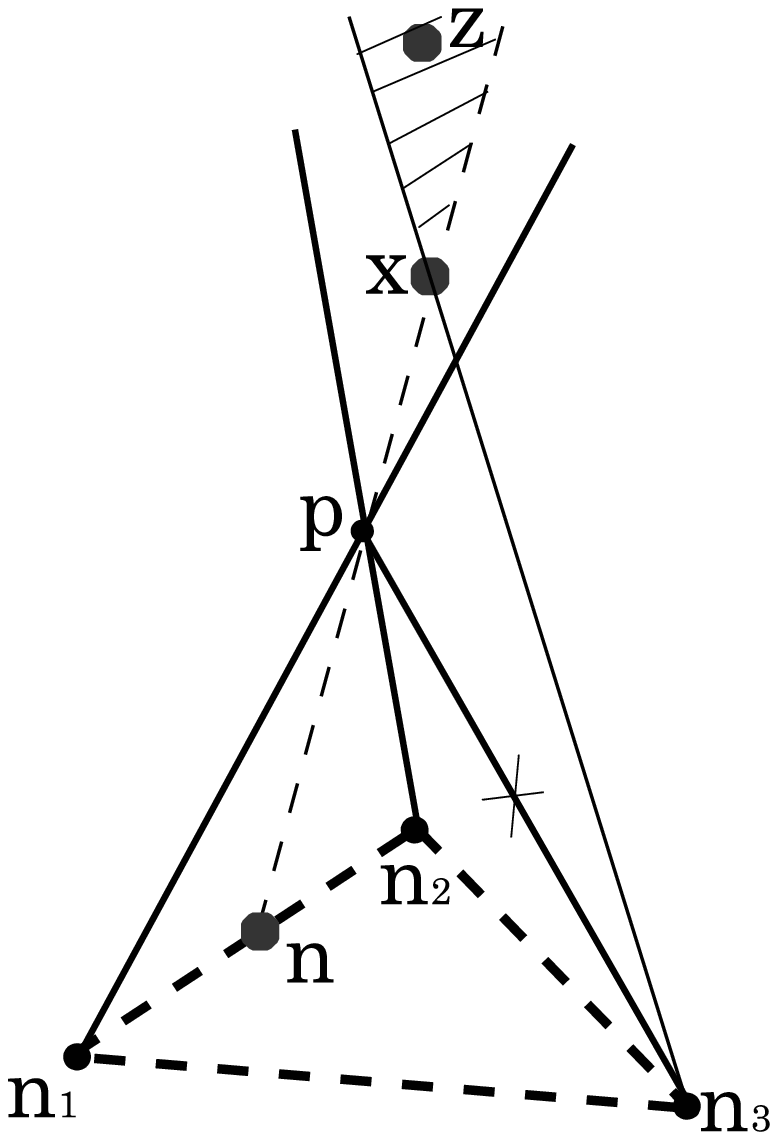} &
\includegraphics[width=3in]{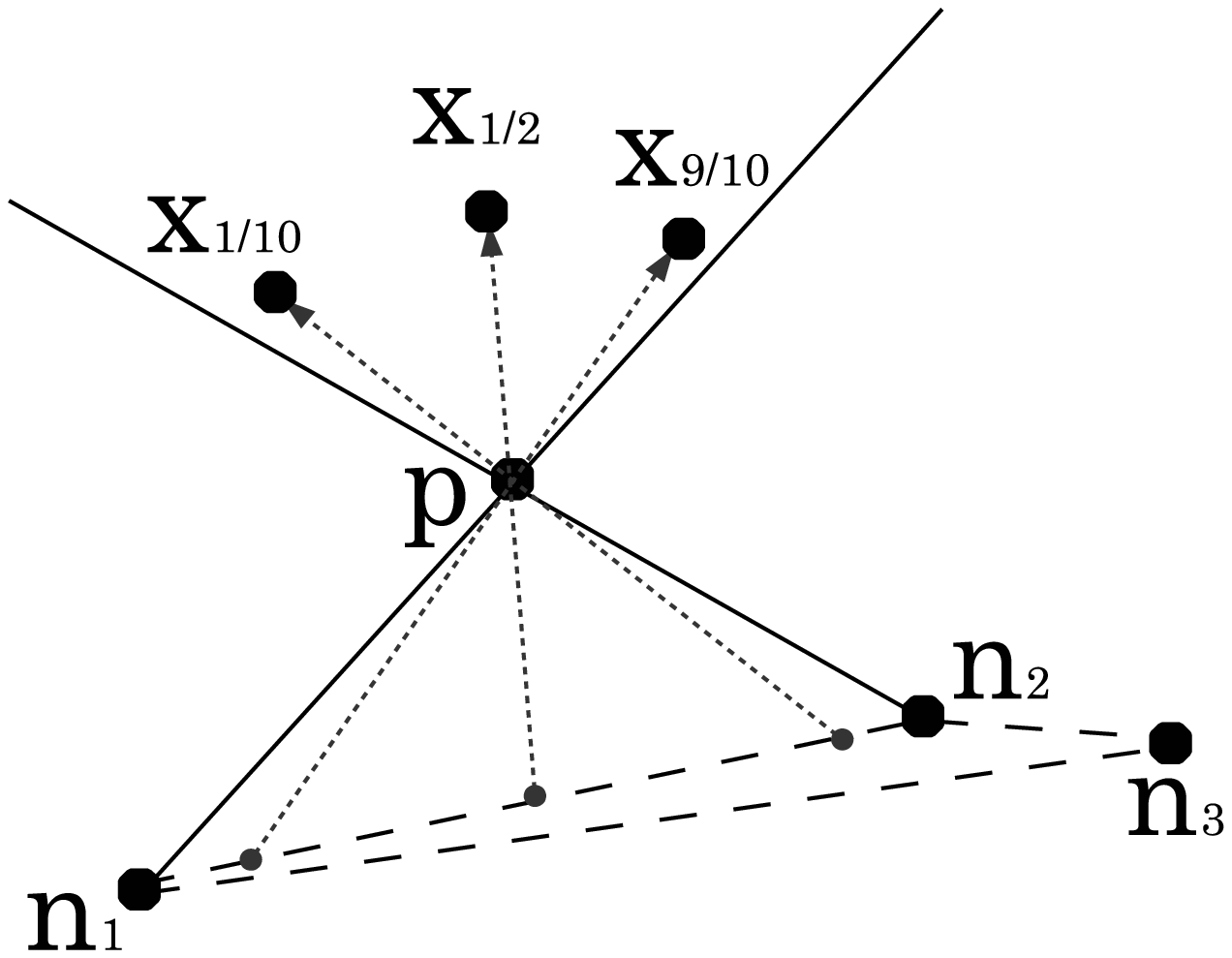}
\end{tabular}
\vspace{-2cm}
\caption{Left: An uncounter-balanced vertex $p$ with at least three neighbors $n_{1}, n_{2}, n{3}$, is counter-balanced by creating bond $\overline{px}$ and deleting bond $\overline{pn_{3}}$. Likewise, $x$ is counter-balanced by finding $z$ such that $x$ is inside triangle $\triangle zpn_{3}$, consequently creating bond $\overline{xz}$. Right: An illustration of different positions to choose an appropriate neighbor $x$ for uncounter-balanced vertex $p$. Values of $\beta$ are chosen so that most of the space contained in the angle determined by vectors $\protect\overrightarrow{pn_{1}}$ and 
$\protect\overrightarrow{pn_{2}}$ is scanned. }
\label{CBpseudocode1}
\end{figure} 

\item \textbf{Counter-balance strategy for a vertex with at least two neighbors:}
Assume the situation depicted in Figure \ref{CBpseudocode2}, where $p$ is an uncounter-balanced vertex with neighbors $n_{1}$ and $n_{2}$. In this case, create a new vertex $x$ such that
	\begin{equation} 
		\overrightarrow{x}=\frac{\overrightarrow{np}}{\|\overrightarrow{pn}\|}\alpha;\quad
		\overrightarrow{n}=\frac{1}{2}\left(\overrightarrow{n_{1}}+\overrightarrow{n_{2}}\right).
	\end{equation}
Then find two connected 
		vertices $a$ and $b$ and create bonds $\overline{xp}$, $\overline{xa}$, $\overline{xb}$ and delete the prior bond 
		$\overline{ab}$. This strategy is successful if the new bonds do \emph{not} cross any of the prior ones and if either vertex  
		$a$ or $b$ remain counter-balanced after deleting the bond connecting them.  It is important to note that one can only apply this strategy for a vertex with just two neighors.   
\end{itemize}

\begin{figure}
\centering
\includegraphics[width=3in]{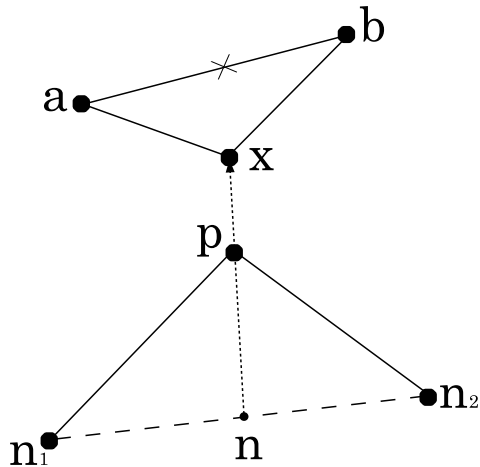}
\vspace{-2cm}
\caption{Uncounter-balanced vertex $p$ is counter-balanced by finding two connected vertices $a$ and $b$ such that the creation of vertex $x$ and bonds $\overline{xp},\overline{xa}$, and $\overline{xb}$ causes $x$ to be counter-balanced. Also, bond $\overline{ab}$ is deleted.}
\label{CBpseudocode2}
\end{figure}

A few comments on the algorithm are in order.  (0) The use of $r_{min}$ and $r_{max}$ set the local neighborhood from which vertices are connected.  It is in this sense that the algorithm is local spatially. (1) We did not impose counter-balance during the Laman construction because Type II Henneberg steps delete bonds so that a counter-balanced vertex at one point during the graph construction may not be counter-balanced at some later point in the construction. Of course, Type I Henneberg steps do not delete bonds, but then a sizeable fraction of uncounter-balanced vertices emerge, approximately 75 percent, as compared to jamming graphs constructed using the Type II Henneberg steps, where approximately 50 percent of the bonds are counter-balanced automatically. (2) The algorithm begins with a triangle, which is a $2N-3$ minimally rigid structure. While periodic boundary conditions imply $N_g=2$, local structures that are the equivalent of $2N-2$ minimally rigid structure do not include single-bonded triangles.  More precisely, for $N=3$, one connection between two vertices must have a double bond. Since we begin the algorithm with a local structure and build up the graph from there with specific boundary conditions imposed for some fraction of connections that decreases as $N$ increases, we implement $N_g=3$.  The $N_g=2$ versus $N_g=3$ for periodic particle packings presumably accounts for why an extra contact is needed a positive bulk modulus. (3) The jamming graph is one connected structure, i.e. there are no rattlers.  Rattlers are usually removed by hand when studying elastic properties, for example, since they do not contribute to the network. (4) New vertices may be added to the jamming graph to enforce the counter-balance property. We use $N$ to denote the number of vertices before counter-balance is enforced such that the approximate number of final vertices is $\frac{3}{2}N$. 

To compare the jamming graph algorithm with other algorithms generating minimally rigid particle packings, such particle-based algorithms range from minimization methods~\cite{ohern2,teitel} to adapative network methods~\cite{tkachenko} to molecular dynamics~\cite{silbert} to event-driven molecular dynamics~\cite{donev}. When using these approaches, the transition point, defined by the packing fraction or the average pressure, can be protocol-dependent~\cite{chaudhuri,vagberg,dagois}. Moreover, convergence issues exist~\cite{goodrich}. Our algorithm uses purely contact geometry to concretely define the onset of jamming in frictionless spheres. Indeed, there exist algorithms to generate generic disordered minimally rigid graphs (no particles) via a matching algorithm~\cite{moukarzel1, moukarzel2}, where generic means that the vertex coordinates are not related by any symmetry. And finally, there exists a hybrid approach where high density particle packings are used to generated a disordered hyperstatic graph~\cite{ellenbroek3}. Then, bonds are then randomly deleted from the graph until, for example, $<z>=4$ is obtained. There is a constraint on the random deletion, however, where a bond is not deleted if the local coordination number of either of the two associated vertices goes below 3, otherwise known as the $k$-core condition~\cite{schwarz1}. However, both latter algorithms do not necessarily enforce the geometry of local mechanical stability.

\section{Perturbing jamming graphs}
Now that we have an algorithm to construct jamming graphs, we perturb these graphs to determine how the system destabilizes (and restabilizes) mechanically. The destabilization is studied via the identification of rigid clusters. A rigid cluster defines those bonds that are rigid with respect to each other. We use the powerful pebble game algorithm ~\cite{jacobs1,jacobs2} to identify rigid clusters via an additional test bond. By construction, the jamming graph is one minimally rigid cluster. We now investigate how the jamming graph destabilizes with the removal of one bond. Note that the size of rigid clusters is measured in terms of bonds and not vertices. 

\subsection{One random bond deletion}
\begin{figure}[h]
\centering
\begin{tabular}{cc}
\includegraphics[width=3in]{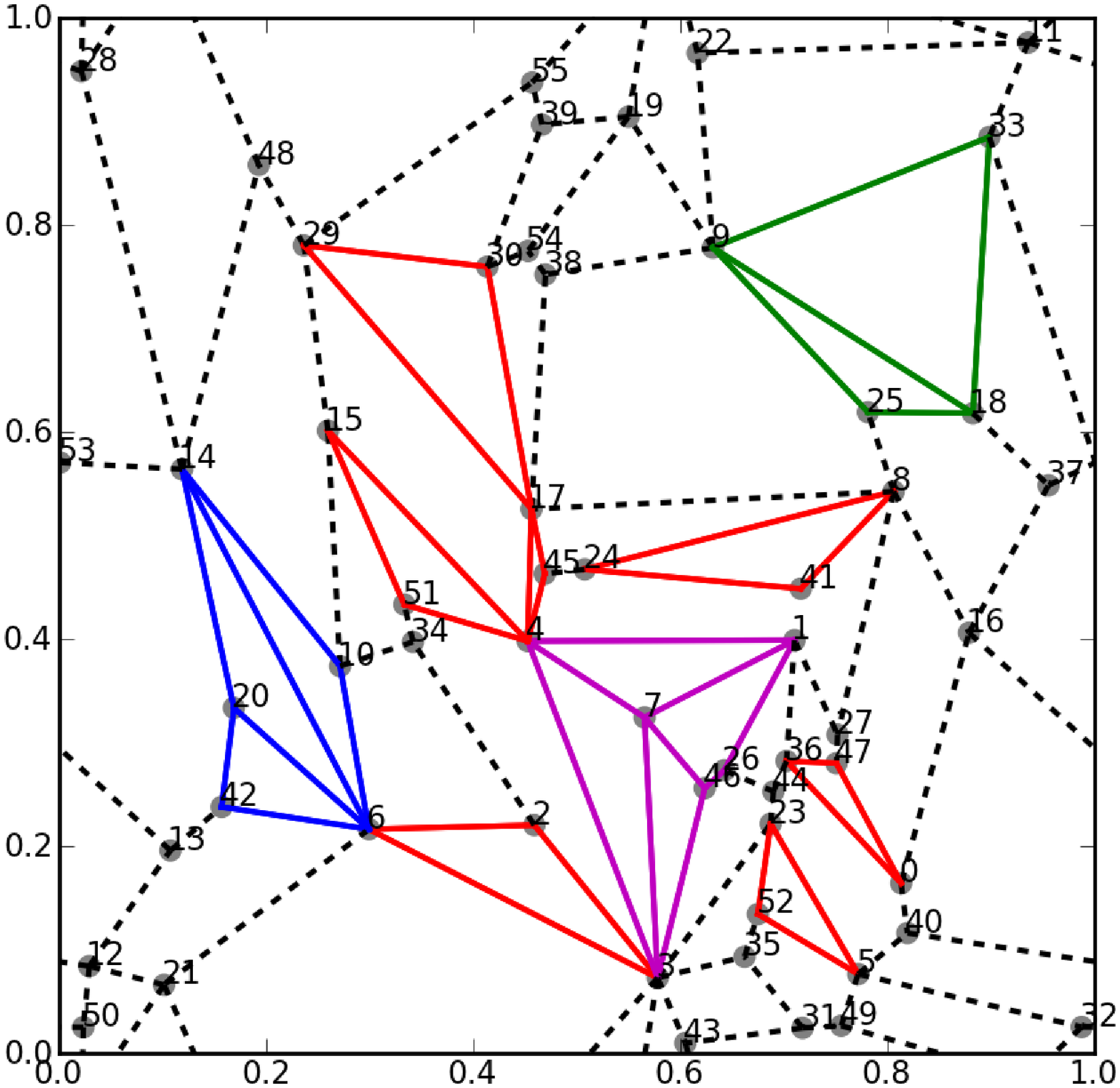} &
\includegraphics[width=3in]{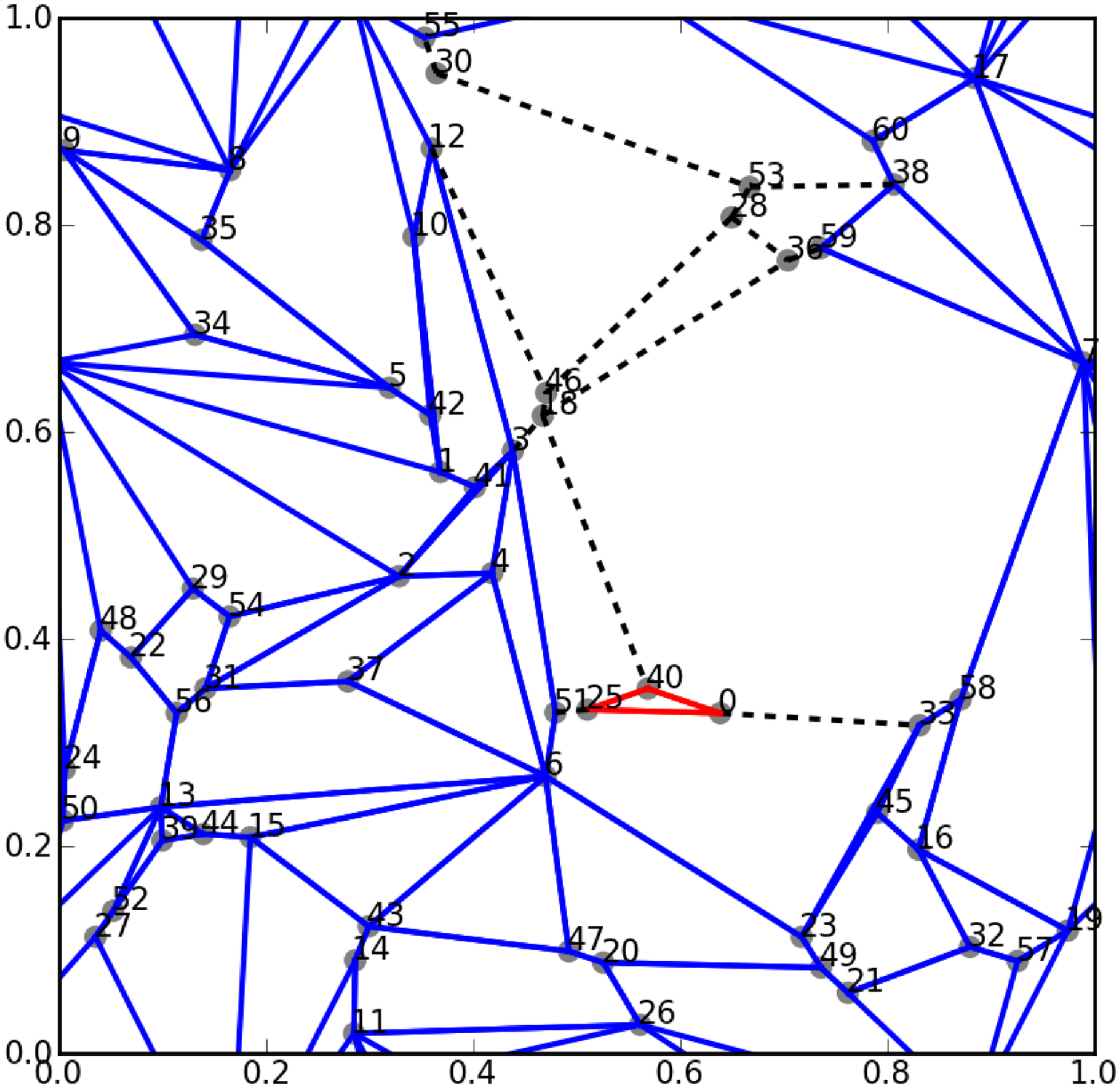}
\end{tabular}
\caption{Left: Many small rigid clusters identified via color after the deletion of one bond/contact, namely the bond between vertices 1 and 41 as numbered. The black dashed lines indicate bonds that are not rigid with respect to any other bond. Right: A ``macroscopic'' rigid cluster along with a few small ones after the deletion of the bond connecting vertices 12 and 30.}
\label{rigidclusterspic}
\end{figure}

The removal of one bond/contact creates exactly one floppy mode such that every bond is no longer rigid with respect to every other bond in the graph. In other words, there must be at least two rigid clusters in the graph. In fact, one can prove that there must be an even number of rigid clusters. So, precisely how many pairs of rigid clusters are there after one bond is randomly deleted?  Many microscopic rigid clusters with no macroscopic one, or at least one macroscopic rigid cluster coexisting with microscopic sized rigid clusters? We define macroscopic rigid cluster to occupt some finite fraction of the bonds in the system and scale with the system size.  Microscopic rigid clusters, on the other hand, do not scale with the number of bonds in the system  The smallest minimally rigid cluster is a triangle. It turns out that both scenarios are observed. See Figure ~\ref{rigidclusterspic}. And while both scenarios occur, it turns out that the most common scenario is the absence of at least one macroscopic rigid cluster when one bond is deleted.

\begin{figure}
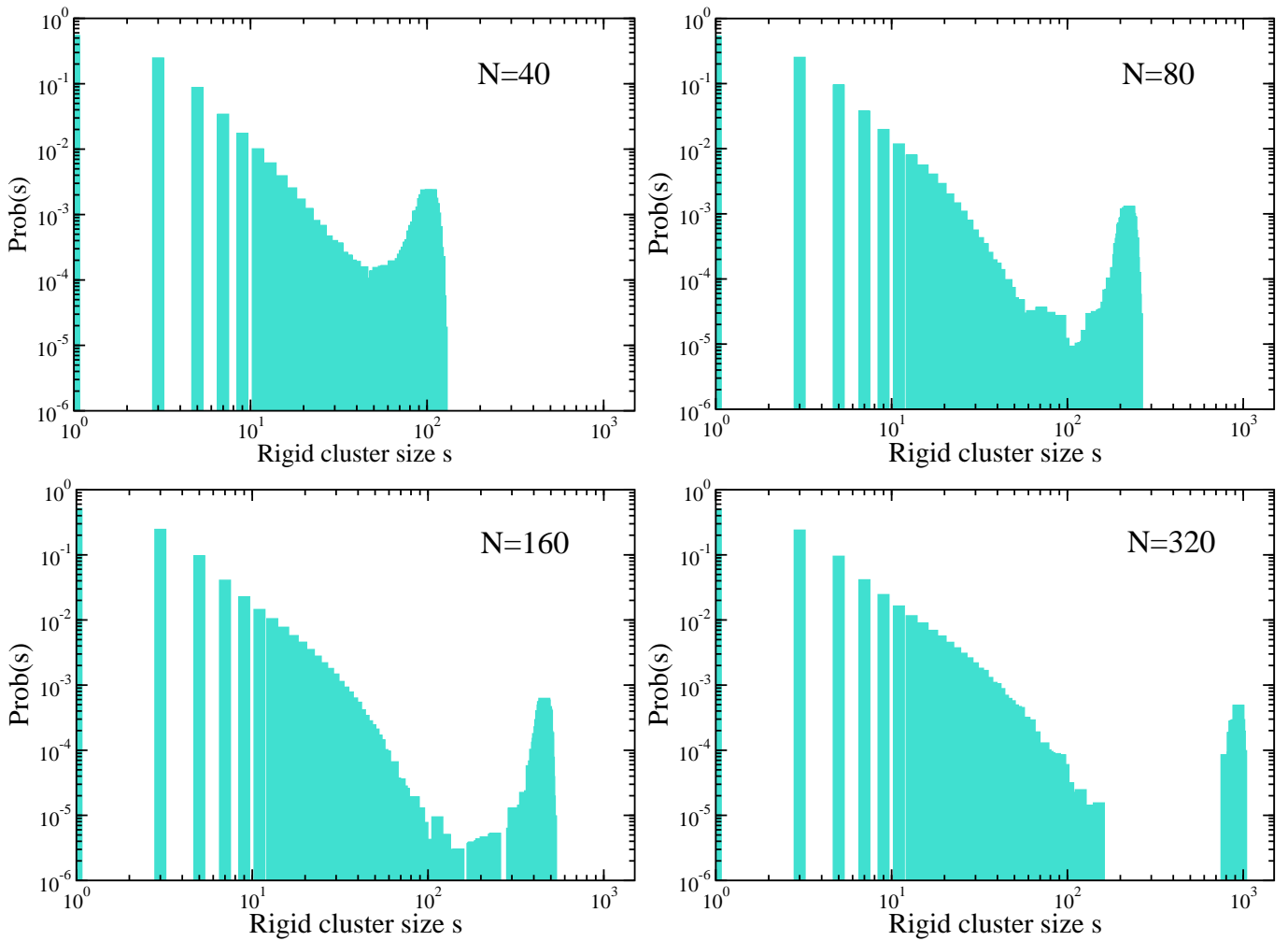

\centering
\begin{tabular}{cc}
\epsfig{file=isostatic.N40.del1.100000samples.eps,width=0.5\linewidth,clip=} &
\epsfig{file=isostatic.N80.del1.100000samples.eps,width=0.5\linewidth,clip=} \\
\epsfig{file=isostatic.N160.del1.100000samples.eps,width=0.5\linewidth,clip=} &
\epsfig{file=isostatic.N320.del1.10000samples.eps,width=0.5\linewidth,clip=}
\end{tabular}
\caption{Plot of $Prob(s)$, the probability for a bond to participate in a rigid cluster of size $s$ after one bond is deleted from the jamming graph. The different graphs represent different system sizes.}
\label{rigidclusterhistogram}
\end{figure}

\begin{figure}[h]
\centering
\includegraphics[width=3in]{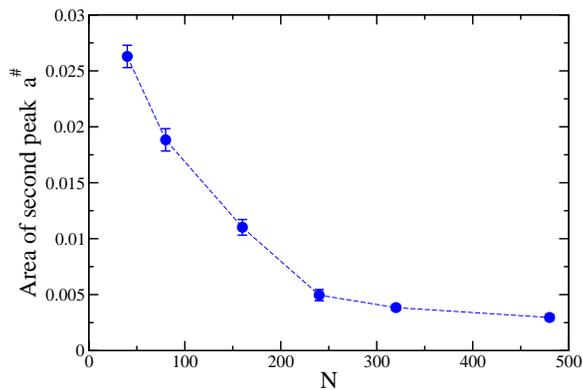}
\caption{Plot of the peak area for the macroscopic rigid cluster sizes, $a^\#$, as a function of $N$ initial vertices. The dashed line is merely a guide for th eye.}
\label{peakarea}
\end{figure}

And while the survival of macroscopic rigid clusters after one bond deletion is apparent in the systems studied, do they persist in the large system limit? Figure~\ref{rigidclusterhistogram} depicts the resulting rigid cluster size probability distribution, $Prob(s)$ for $N=40$, $80$, $160$, and $320$.  Indeed, the probability of observing a macroscopic rigid cluster after one bond deletion decreases with increasing system size. To obtain a systematic measurement, we compute the area of the characteristic peak at the macroscopic rigid cluster sizes $s$. See Figure~\ref{peakarea}. While the trend is not clearly power-law nor exponential, the area, $a^\#$, is decreasing with $N$, suggesting that macroscopic rigid clusters after one bond deletion vanish in the infinite system limit.  This possibility, while not as likely as the many, microscopic rigid clusters scenario, cannot yet be completely ruled out in the infinite system limit, i.e. there is no mathematical proof. 

\begin{figure}[h]
\centering
\includegraphics[width=2in]{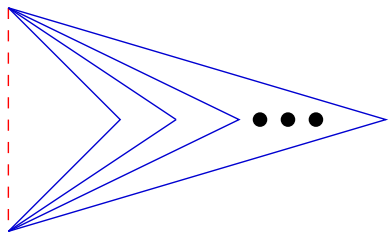}
\caption{Example graph where removal of the red, dashed line leads to all remaining bonds not rigid with respect to each other. The removal of any one blue line affects only its neighboring blue bond.  The rest of the graph is unaffected. The three black circles denote repetition of the blue hinge bonds.}
\label{example}
\end{figure}

For the sake of argument, consider the graph in Figure ~\ref{example}~\cite{servatius}.  Removal of the red dashed bond corresponds to each bond not being rigid with respect to any other, or many rigid clusters of size one.  However, removal of one of the blue bonds leaves the rest of the rigid structure unchanged (except for the one neighboring bond that is no longer rigid with respect to any other bond). Depending on the bond that is deleted, either scenario holds in the infinite system limit. And while the specificity of this graph may not be useful for understanding the generic case, extensions of such examples may indeed be.

In the system sizes studied, the two scenarios---(1) many microscopic rigid clusters with no macroscopic rigid cluster and (2) at least one macroscopic, rigid cluster---can be related to two different floppy modes. In the many microscopic rigid clusters scenario, there is widespread break-up of the system.  For bonds between individual rigid clusters, there is zero-energy cost to deforming those bonds such that if they are replaced by springs as is typically done when determining vibrational modes, these bonds contribute to any zero-energy modes.  If these bonds extend across the system, as they do in the case of many microscopic rigid clusters, then the zero-energy mode is an extended one. In contrast, the presence of at least one macroscopic rigid cluster translates to a more localized zero-mode since the deforming bonds within the macroscopic rigid cluster will result in some energy cost. 

Our finding is presumably related to the recent finding of two kinds of instabilities due to the breaking of contacts in a repulsive, frictionless particle packing at the onset of rigidity~\cite{lerner}. In this work, two particles in contact are pulled apart, i.e. a bond is deleted, and particle rearrangements driven by instabilities are identified. There appear to be two kinds of instabilities---one extended and the other localized. While force information is assumed in this study, our work only uses contact information.

In addition to the decrease in $a^\#$ with increasing $N$, there is another trend in the probability distribution of rigid cluster sizes as the system size increases. There is a suppression of intermediate cluster sizes starts to emerge, i.e. a gap between the microscopic and macroscopic rigid clusters emerges.  There exists an upper bound on the small rigid clusters that does not change with increasing system size. Interestingly, Ref.~\cite{goodrich} argued for the absence of intermediate rigid cluster sizes when introducing a surface of cut bonds into the system based on a necessary, but not sufficient, condition for rigidity. 

What does a system size-independent upper bound on the microscopic rigid cluster sizes imply about length scales?  Should the macroscopic rigid cluster scenario vanish in the infinite system limit, a diverging lengthscale emerges in the sense that going from one macroscopic rigid cluster to many microscopic rigid clusters in the infinite system limit corresponds to an infinite length associated with catastrophic break-up of the one minimally rigid cluster.  

\begin{figure}
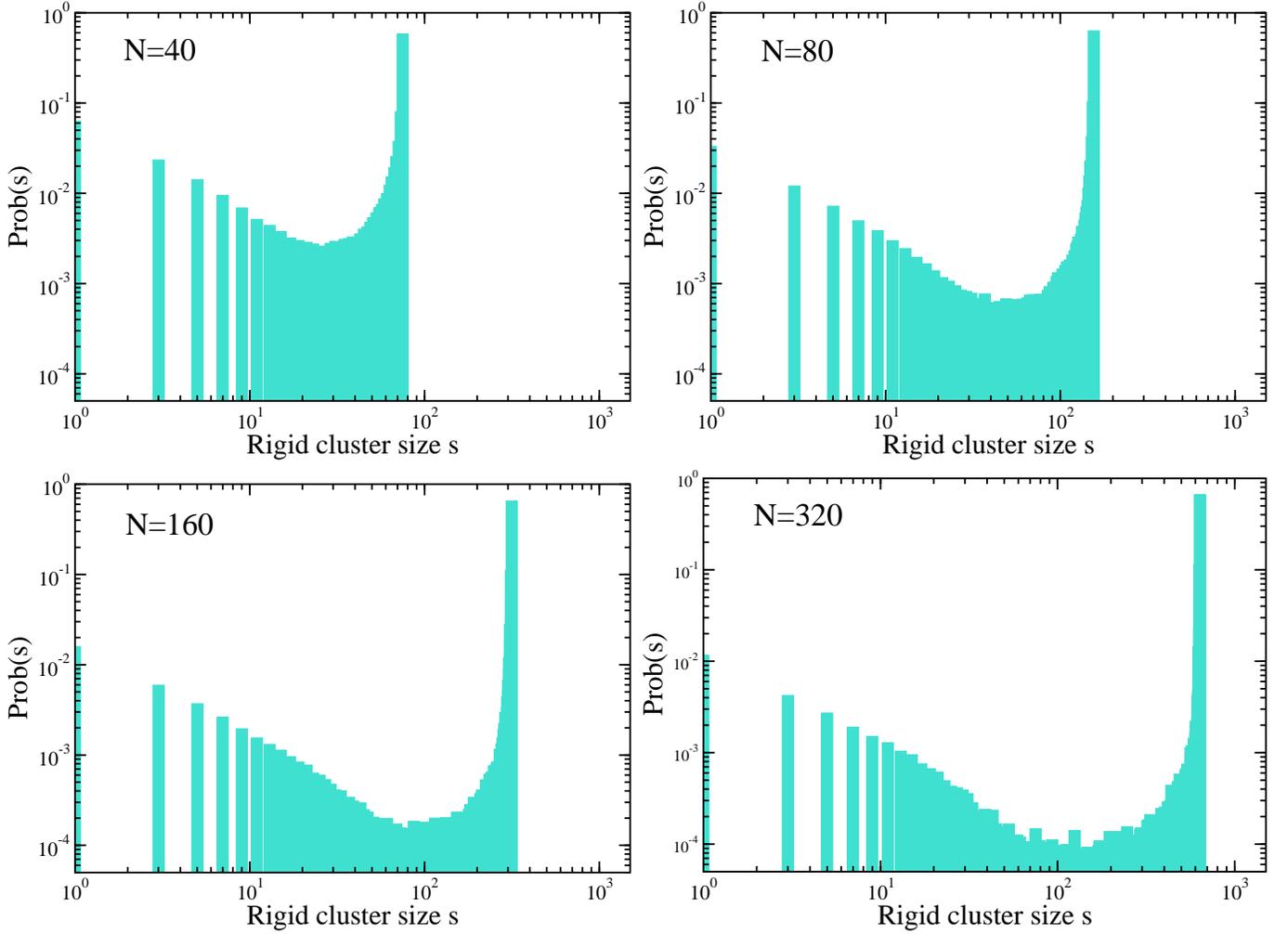

\centering
\begin{tabular}{cc}
\epsfig{file=isostatic.type1.N40.del1.100000samples.eps,width=0.5\linewidth,clip=} &
\epsfig{file=isostatic.type1.N80.del1.100000samples.eps,width=0.5\linewidth,clip=} \\
\epsfig{file=isostatic.type1.N160.del1.100000samples.eps,width=0.5\linewidth,clip=} &
\epsfig{file=isostatic.type1.N320.del1.25000samples.eps,width=0.5\linewidth,clip=}
\end{tabular}
\caption{Plot of $Prob(s)$, the probability for a bond to participate in a rigid cluster of size $s$ after one bond is deleted from the minimally rigid graphs generated by Henneberg Type I moves only. The different graphs represent different system sizes.}
\label{type1}
\end{figure}

To compare our results for the jamming graph with other minimally rigid graphs, ones where local mechanical stability need not be obeyed, we generated minimally rigid graphs using the Henneberg Type I and did not enforce local mechanical stability (counter-balance) for each vertex.   We then delete one bond randomly and compute the resulting rigid cluster probability distribution.  See Figure ~\ref{type1}. As with jamming graphs, the resulting rigid cluster distribution exhibits two scenarios---one with many microscopic rigid clusters (and no macroscopic rigid cluster) and another less typical scenario with at least one macroscopic rigid cluster---demonstrating a similar trend to the jamming graph case. These results, again, suggest that there are both extended and localized zero-energy modes in a minimally rigid system. For the Type I graphs, the gap between the macroscopic and microscopic rigid clusters does {\it not} emerge as clearly as compared to the jamming graph at similar system sizes. 

\begin{figure}
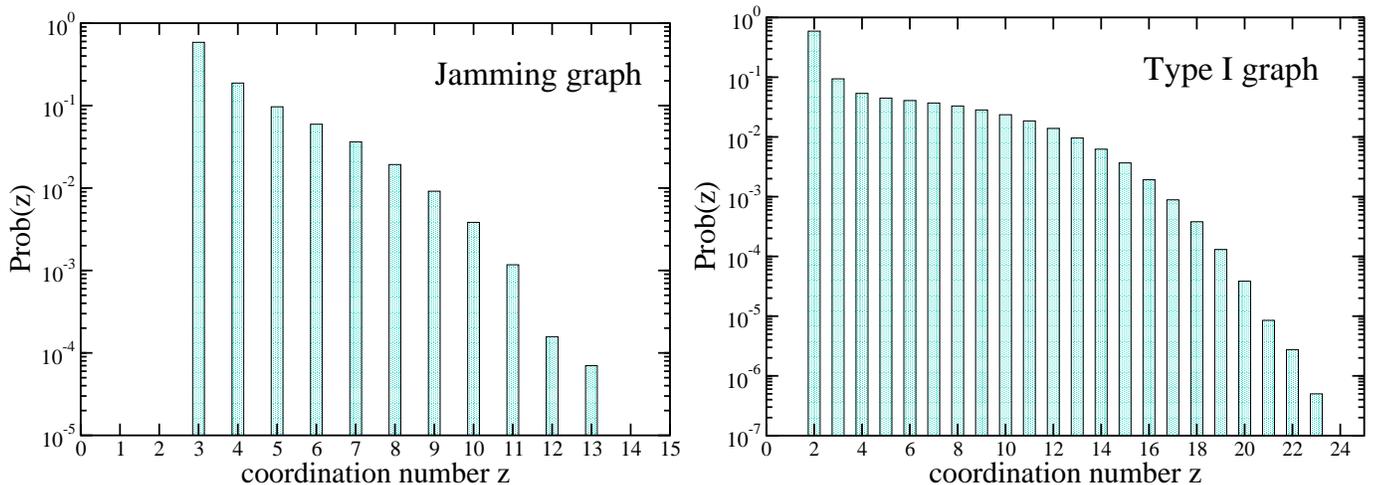

\centering
\begin{tabular}{cc}
\epsfig{file=connectivity.N40.eps,width=0.5\linewidth,clip=} &
\epsfig{file=connectivity.type1.N40.eps,width=0.5\linewidth,clip=} \\
\end{tabular}
\caption{Plot of $Prob(z)$, the probability for a site to have $z$ neighbors, for the jamming graphs (left) and the graphs generated using Henneberg Type I moves (right). For both plots, $N=40$.}
\label{connect}
\end{figure}

This difference could be due to different connectivity of the Type I graphs, which is less constrained than the jamming graphs.  See Figure ~\ref{connect}. In particular, the fraction of vertices with just two bonds is about 59 percent. The removal of either bond removes the possibility of that particular vertex participating in any rigid cluster of at least size three. While for the jamming graph, the removal of a bond does not prevent the neighboring vertex of now two or more neighbors from taking part in another local rigid cluster. In other words, the rigid cluster structures are not as local as in the jamming graph such that one may have to go to much larger system sizes to observe a gap in the rigid cluster sizes.  Interestingly, Goodrich and collaborators~\cite{goodrich} did not observe a sudden loss of rigidity for a subsystem size below some lengthscale for bond-diluted hexagonal lattices where vertices of coordination number two are allowed. As with Type I graphs, it may be that one must go to much larger system sizes to see a gap emerge.

After deleting one bond, let us briefly discuss how the system restabilizes mechanically (or not) with the random addition of one bond. If the added bond is within one rigid cluster, then the bond is redundant and the graph does not restabilize mechanically. However, if the added bond is between two rigid clusters, then it is an independent constraint such that the graph becomes minimally rigid and, therefore, must restabilize mechanically, i.e. there is only one rigid cluster. In terms of a lengthscale affected by the perturbation, if the destabilized system is made up of many microscopic rigid clusters, the lengthscale affected by the perturbation is of order the system size, i.e. it is infinite in the infinite system limit, since the system goes from many microscopic rigid clusters to one macroscopic rigid cluster with the addition of one bond.

\subsection{Hyperstatic: Randomly adding and deleting more than one bond}

We begin with a jamming graph and randomly add some number of bonds, $A$. The graph is no longer minimally rigid, i.e. it is now hyperstatic. This action allows us to study systems that are ``above'' jamming where the macroscopic rigid clusters survive at least one bond deletion. We can then randomly delete (different) bonds from this hyperstatic graph and investigate how the system destabilizes mechanically. We may then be able to identify a lengthscale that decreases from the system size at jamming to some lengthscale smaller than the system size as a result of randomly deleting $D$ bonds.  

\begin{figure}
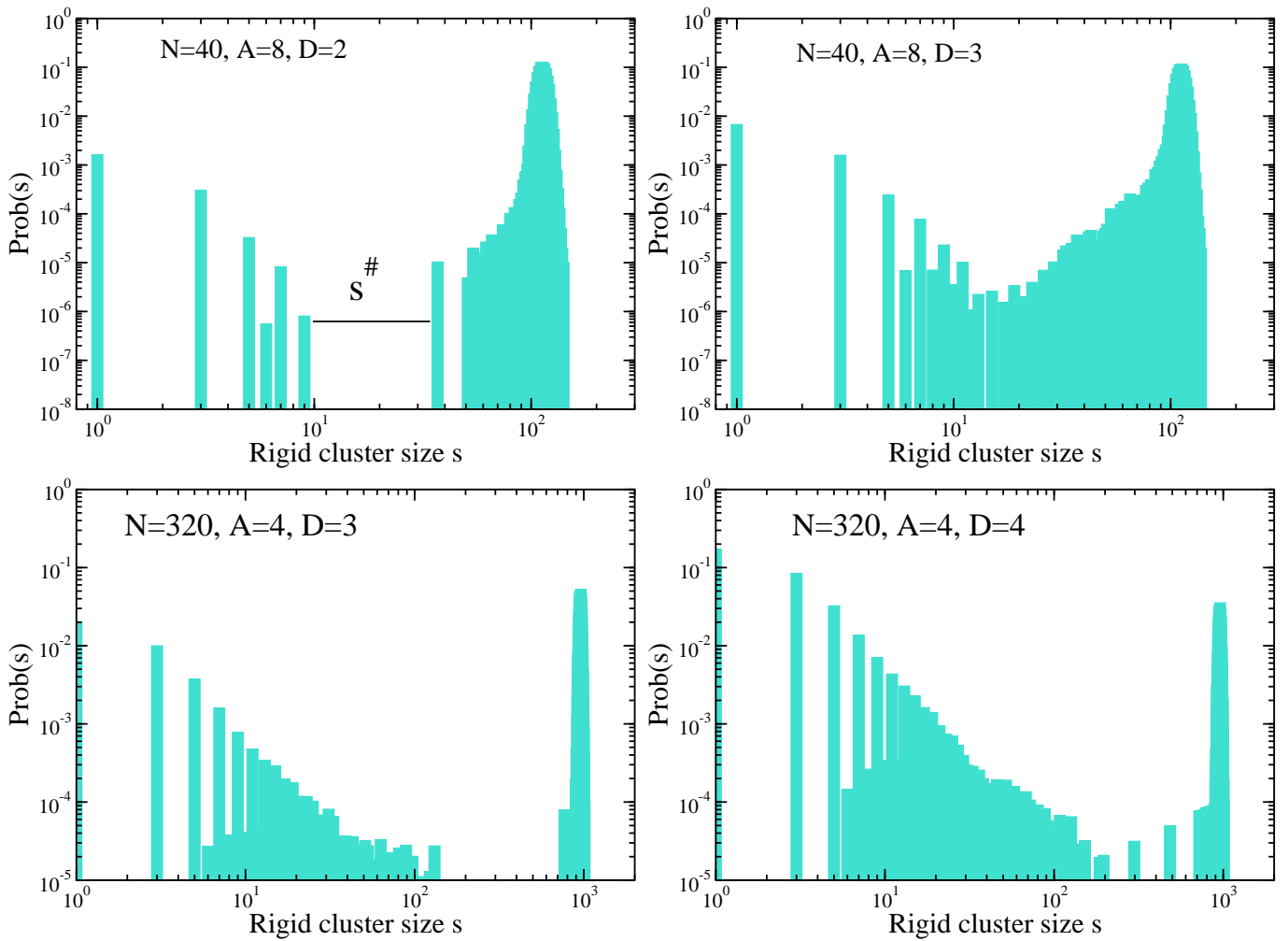

\centering
\begin{tabular}{cc}
\epsfig{file=hyperstatic.N40.add8.del2.eps,width=0.5\linewidth,clip=} &
\epsfig{file=hyperstatic.N40.add8.del3.eps,width=0.5\linewidth,clip=} \\
\epsfig{file=hyperstatic.N320.add4.del3.eps,width=0.5\linewidth,clip=} &
\epsfig{file=hyperstatic.N320.add4.del4.eps,width=0.5\linewidth,clip=}
\end{tabular}
\caption{Plot of $Prob(s)$, the probability for a bond to participate in a rigid cluster of size $s$ after $A$ bonds are randomly added to the jamming graph and then $D$ bonds are randomly removed.  Two different system sizes are shown.}
\label{hyperstatic}
\end{figure}

For a concrete example, consider adding eight redundant bonds to a jamming graph with $N=40$. For 100,000 realizations, the random removal of one bond does not create more than one rigid cluster, i.e. the system is still rigid globally. The random removal of two bonds creates a few small rigid clusters in addition to the macroscopic rigid cluster with a gap in between. See Figure ~\ref{hyperstatic}. When three bonds are randomly removed, however, we observe more of a qualitative change in the rigid cluster size distribution.  The gap between microscopic rigid clusters and macroscopic rigid clusters closes.  The concept of a single system with microscopic rigid clusters separated macroscopic rigid clusters no longer makes sense. We are now beginning to observe extended break up of the system. The gap size just before the extended break-up sets a size scale, $s^\#$. This size scale can be easily converted to a lengthscale via $s^\#\sim (l^{\#})^2$ in two-dimensions (assuming compactness). Note that as the system size becomes larger such that one can approach the transition more closely, the extended break up of the system results in the absence of any macroscopic rigid clusters. 

How does $s^\#$ scale with $N$ and with distance to the rigidity transition? We define $\epsilon=\frac{A}{N}$ to describe the distance to the transition. More specifically, since $<z>=\frac{2}{N_f}(2N_f-3+A)$, with $N_f$ representing the final number of vertices in the graph xafter counter-balancing, then
\begin{equation}
<z>-4+\frac{3}{N_f}=\frac{4}{3}\epsilon,
\end{equation} 
where we have included the $1/N_f$ correction to the location of the transition~\cite{laman,goodrich2} and used $\frac{N}{N_f}=\frac{2}{3}$. For fixed $\epsilon$, we observe that $s^\#$ increases with increasing $N$ until beginning to reach a plateau that is independent of system size. We also observe that as $\epsilon$ decreases, $s^\#$ increases, though it will ultimately be confined by the system size.  In other words, $s^\#$ is behaving as a size scale near a critical point. See Figure 14.

To test this hypothesis, we measure $s^\#$ as a function of $\epsilon$ for several system sizes and attempt finite-size scaling via the following route. Let $l\sim \epsilon^{-\nu}$ and $N\sim l^2$, where $l$ is some underlying diverging lengthscale in the system. If the observed diverging lengthscale, $l^\#$, is due to the underlying diverging lengthscale, then $l^\#\sim \epsilon^{-\rho}$, again, with $s^\#\sim (l^{\#})^2$ such that 
\begin{equation}
l^\#=l^{\rho/\nu}f(l^{1/\nu}\epsilon),
\end{equation}
with $f(y)$ as some universal scaling function. For the cut-out subsystem analysis with either free or fixed boundary conditions, $\nu=1$. If we assume this and set $\rho=\nu$, then we do not obtain a good scaling collapse. A more recent discovered diverging lengthscale, $l_c$, associated with the localization of phonon modes in floppy networks, results in $\nu=1/2$~\cite{during,wyart3}. If we, again, assume $\rho=\nu$, we do not obtain a good scaling collapse.  We also tried the remaining the two remaining combinations of $\nu=1,\,\rho=1/2$ and $\nu=1/2,\,\rho=1$ and did not obtain good collapse. However, with $\nu=1/3$ and $\rho=1/2$, we do obtain a good scaling collapse.  See Figure 14. While this data is suggestive of perhaps a new diverging lengthscale in the system associated with point perturbations via $\nu=1/3$, this possibility may be ruled out by the study of larger system sizes. 

It is interesting to note that mean-field correlation length exponents of $1/4$, $1/2$, and $1$ exist in disordered systems~\cite{schwarz1,wyart1,wyart2,during,wyart3}, while a correlation length exponent of $1/3$ is presumably less common. Incidentally, while $\nu=1/4$ (and $\rho=1/2$) is not as good a collapse as with $\nu=1/3$, it cannot be ruled out at this stage. However, we can conclude that $\nu=1,1/2$ and/or $\rho=1,1/2$ does not lead to sufficient collapse for at least the system sizes we study.

\begin{figure}
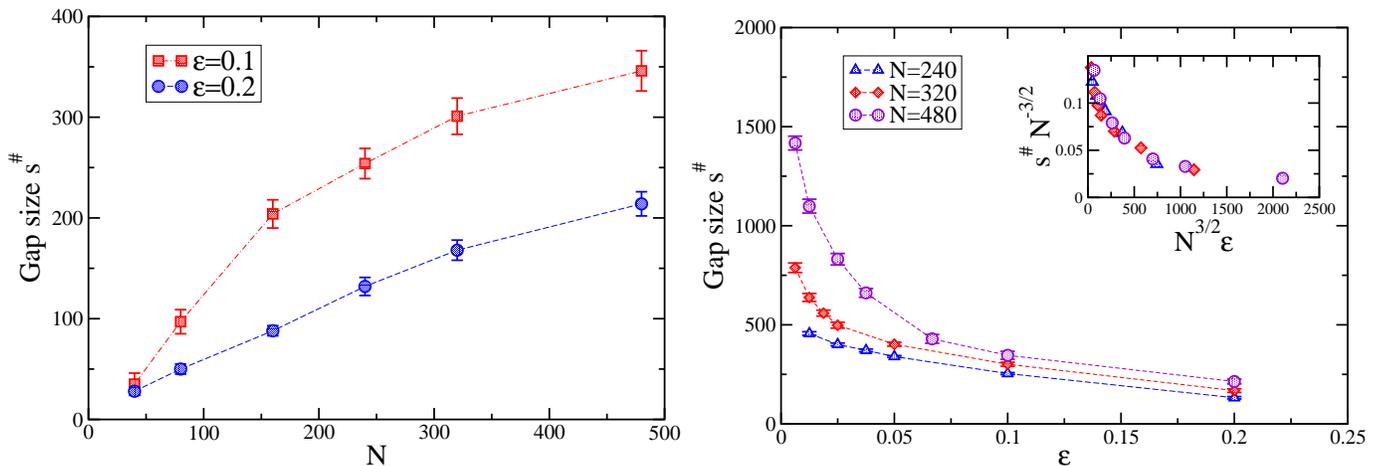

\centering
\begin{tabular}{cc}
\epsfig{file=gapsize.versusN.epsilon.eps,width=0.5\linewidth,clip=} &
\epsfig{file=gapsize.versusA.collapse.eps,width=0.5\linewidth,clip=} 
\end{tabular}
\caption{Left: Plot of the gap size $s^\#$ as a function of $N$ for fixed 
$\epsilon=\frac{A}{N}$. Right: Plot of $s^\#$ as a function of $\epsilon$ for different system sizes.  The inset is a finite-scaling scaling collapse obtained with $\nu=\frac{1}{3}$ and $\rho=\frac{1}{2}$. All dashed lines are guides for the eye.}
\label{gapsize}
\end{figure}

\section{Discussion}

We have presented an algorithm for the spatially local construction of jamming graphs. Jamming graphs represent the contact geometry of soft repulsive, frictionless discs at the onset of mechanical rigidity. In other words, they contain both the property of local mechanical stability and the necessary and sufficient condition for minimal rigidity via the Henneberg construction.  Since the bonds represent contacts between particles/vertices, the jamming graph is planar. Varying the construction of these (and related) graphs will allow us to turn on/off different properties of particle packings at/near the onset of rigidity in a controlled way to form a more concrete and comprehensive picture.

Our construction of jamming graphs begs at least three immediate extensions of study. For the first extension, if one associates a disc with each vertex and each bond dictates the contacts between discs, then we can potentially analyze the question of whether or not there exists a unique packing fraction at the onset of rigidity in the infinite system limit~\cite{torquato,ohern2,zamponi1,chaudhuri,zamponi2}. For example, one could define the onset of rigidity to occur at the random closed packing (RCP) point~\cite{rcp,ohern2}, or a maximally jammed state~\cite{torquato}, or a jamming line~\cite{zamponi1,zamponi2}. To address this issue, we can create a physical realization of discs from a jamming graph using the circle packing theorem~\cite{CirclePacking}. The circle packing theorem states that for every connected simple planar graph $G$ there exists a circle packing in the plane whose intersection graph is isomorphic to $G$. There are more strict conditions for uniqueness as such and these will be explored.  While we can immediately explore this issue in two-dimensions where both Laman's theorem is exact and the circle packing theorem holds, it would be interesting to extend the jamming graph to three-dimensions and higher. It turns out that the three-dimensional version of Laman's theorem for some networks are essentially exact~\cite{thorpe}. Higher-dimensional extensions of the circle packing theorem would indeed be more challenging. 

As for the second immediate extension, with jamming graphs we can search for interplay between global mechanical stability and local mechanical stability since we can, for example, easily turn off/on local mechanical stability. This on/off switch allows us to compare the mechanics of fixed connectivity networks with repulsive particle packings~\cite{ellenbroek3,during,tighe1,tighe2}.  While local mechanical stability may not play as much of a role right at the transition (other than suppressing fluctuations~\cite{wyart2}), it will certainly play a role in particle rearrangements above jamming where local mechanical stability is also required. For instance, if particle chains form as a result of some perturbation, the chain should buckle so that particles having two contacts will ultimately have at least three contacts. Such particle rearrangements above jamming can also be studied using jamming graphs with the removal and addition of bonds. For instance, if the breaking of a contact results in a vertex not being counter-balanced, with some input of force information~\cite{ellenbroek4,tighe3,xu,lerner}, a sequence of moves can be generated to regain the local mechanical stability while ensuring that the graph remain hyperstatic.  

For the third immediate extension, local rules governing a system may lead to a field theory, should one exist. And while there is not necessarily consensus on a field theory for jamming~\cite{henkes1,zamponi3}, our spatially local construction of the jamming graph may provide insight for building an alternative field theory for a system with both local and global constraints. Such a framework may accelerate the quest to determine how to classify the various constraints in terms of potentially different universality classes. For instance, enforcing only the local $k$-core constraint leads to one type of phase transition, while the counter-balance constraint leads to another~\cite{schwarz1,schwarz2}. And while such attempts are not currently appreciated, a classification system based on constraints will ultimately emerge.  

For instance, the notion of constraints changes when one deviates from soft repulsive, frictionless discs. Ellipsoidal particle packings may or may not be isostatic at the onset of rigidity depending on which degrees of freedom can be accessed~\cite{ellipse1,ellipse2}. It turns out that one can also extend Laman's theorem in two-dimensions to systems with other degrees of freedom~\cite{ileana} and, correspondingly, extend the Henneberg construction. Such an extension of the jamming graph may, therefore, prove useful for understanding the onset of rigidity for nonspherical particles.  As for frictional systems, while the history of the contact information may be difficult to incorporate into an equivalent jamming graph, one can extend the pebble game (to a (3,3) pebble game) to map out the rigid clusters of frictional packings to compare with frictionless packings ~\cite{henkes2}.  These endeavors (and others) will help form a concrete framework for the onset of rigidity in disordered systems. 

After constructing these jamming graphs, we perturbed them by removing one bond to study how the system destabilizes. In the system sizes studied, there exist two scenarios, one where the removal of the bond leads to catastrophic collapse of the macroscopic rigid cluster (an extended zero-energy mode) and the other where the macroscopic rigid cluster suvives (a localized zero-energy mode). As the system size increases, the probability of the localized zero-energy mode decreases.  It would be interesting to prove whether or not this probability vanishes in the infinite system limit. Particularly in two-dimensions, there is a wealth of mathematical literature on minimal rigidity to potentially go beyond heuristic arguments and numerics~\cite{wyart1}.      

As opposed to uncovering a diverging lengthscale in surface versus bulk effects~\cite{wyart1,wyart2,tighe1,mailman1,goodrich}, we have potentially uncovered a new diverging lengthscale in the rigid phase in response to random bond deletion. With a correlation length exponent close to $1/3$, it appears that this new lengthscale is not related with the introduction of a force monopole in the particle packings~\cite{ellenbroek2}. However, once forces are introduced and/or the contact geometry is allowed to rearrange as the system responds to the point perturbation, then one should presumably obtain the prior result. Again, the ability to build upon the jamming graph by incorporating further detail bit-by-bit will allow us to identify the properties ultimately dictating a particular behavior or response.

The authors acknowledge helpful discussion with Silke Henkes, Xu Ma, and Brigitte Servatius. JMS acknowledges funding from NSF-DMR-CAREER-0607454 and the Aspen Center of Physics where part of this work was completed.


\begin{thebibliography}{500}
\bibitem{servatius} This example was pointed out to us by Brigitte Servatius.
\end{thebibliography}
\end{document}